\documentclass[prx,reprint,showpacs]{revtex4-1}
\usepackage{amsmath}
\usepackage{amsfonts}
\usepackage{revsymb}
\usepackage{graphicx}
\usepackage[usenames, dvipsnames]{xcolor}
\usepackage[normalem]{ulem}

\def\ket#1{|\,#1 \,\rangle}
\def\bra#1{\langle \, #1 \,|}
\def\ev#1{\langle #1 \rangle}

\def\eref#1{Eq.~\ref{#1}}

\definecolor{gleb}{rgb}{1,0.5,0}

\definecolor{jaren}{rgb}{0,0.5,1.0}

\newcommand\redout{\bgroup\markoverwith{\textcolor{red}{\rule[.5ex]{2pt}{0.4pt}}}\ULon}

\begin{document}
\title{Oscillating magnetic field effects in high precision metrology}
\author{H. C. J. Gan} 
\author{G. Maslennikov}
\author{K.-W. Tseng}
\author{T. R. Tan}
\author{R. Kaewuam}
\author{K. J.  Arnold}
\author{D. Matsukevich}
\author{M. D. Barrett}
\email{phybmd@nus.edu.sg}
\affiliation{Center for Quantum Technologies, 3 Science Drive 2, Singapore, 117543}
\affiliation{Department of Physics, National University of Singapore, 2 Science Drive 3, Singapore, 117551}
\begin{abstract}
We examine a range of effects arising from ac magnetic fields in high precision metrology.  These results are directly relevant to high precision measurements, and accuracy assessments for state-of-the-art optical clocks.  Strategies to characterize these effects are discussed and a simple technique to accurately determine trap-induced ac magnetic fields in a linear Paul trap is demonstrated using $^{171}\mathrm{Yb}^+$.
\end{abstract}
\maketitle
An ion trap is a widely used tool in atomic physics and a cornerstone system in high precision metrology.  The key advantages of the ion trap system are the high degree of control of individual ions and the rigorous assessment of systematic effects from the environment including the trapping apparatus itself.  In a linear Paul trap, effects that have not been given sufficient attention are those due to magnetic fields arising from trap-induced rf currents in the electrodes.  To our knowledge there are very few instances in which these fields have actually been measured or at least an attempt made to quantify their influence on experiments \cite{berkeland2002laser, AlIon1, lewty2013experimental, meir2018experimental, hoffman2013radio}.  When values were quantifiable they were typically a few $\mathrm{\mu T}$.  Such values would have a significant contribution to many error budgets in high precision metrology.

The primary effect of ac magnetic fields is to shift atomic energy levels.  However, as observed in \cite{meir2018experimental}, they can also influence the assessment of micromotion, which can have further consequences to the validity or accuracy of an experiment.  High precision measurements can also serve as reference points for other measurements.  Hence we consider it useful to provide a clear description of the effects these fields have, and provide suggestions as to how they might be experimentally assessed.

The paper is divided into two main sections. In the first section the various influences of ac magnetic fields are discussed: specifically, the effect on measured Zeeman splittings and shifts of both microwave and optical clock transitions.  For completeness, a brief discussion on the magnetic blackbody radiation shift is also given. In the second section, two methods to measure the amplitude of an ac magnetic field are discussed. Both methods are sensitive to the orientation of the oscillating field relative to an applied static field.  The discussion is focussed mainly on ion trap systems, but the effects are relevant to other time-varying fields, such as line noise, which is also relevant to neutral atom systems.

\section{AC magnetic field effects}
Throughout the rest of the paper, an applied static magnetic field is denoted $B_0$ and its direction is taken as the quantization axis.  The amplitude of an oscillating magnetic field is denoted $B$ and its components orthogonal to and along the quantization axis are denoted $B_\perp$ and $B_z$ respectively.  For any quantity specifying a sensitivity to $\langle B^2 \rangle$, the unit $\mathrm{\mu T}^{-2}$ is in reference to the root-mean-square amplitude of the field.  For quantities specifying a sensitivity to $B$ or one of its components, the unit $\mathrm{\mu T}^{-1}$ is in reference to the amplitude of the applicable field component.

The energy shift of $\ket{a}$ due to an oscillating magnetic field coupled to $\ket{b}$ can be found by direct analogy with an ac stark shift from an oscillating electric field \cite{polarisability}.  With the magnetic dipole operator $\mathbf{M}$ and polarization vector $\mathbf{u}$, the shift is given by
\begin{equation}
\delta E_a=-\frac{\langle B^2\rangle}{2\hbar}\left(\frac{|\bra{b}\mathbf{u}\cdot \mathbf{M}\ket{a}|^2}{\omega_{ba}-\omega}+\frac{|\bra{a}\mathbf{u}\cdot \mathbf{M}\ket{b}|^2}{\omega_{ba}+\omega}\right)
\label{eq1}
\end{equation}
where $\langle \cdot \rangle$ denotes time averaging and $\omega_{ba}=\omega_b-\omega_a$.  This expression is simply the magnetic counterpart of the expression for an ac stark shift from an oscillating electric field \cite{polarisability}.  When $\omega \ll |\omega_{ba}|$ and $\mathbf{u}=\mathbf{e}_0$, this expression reduces to the static quadratic Zeeman shift of $\ket{a}$ due to the magnetic coupling to $\ket{b}$.  In this case, the effect of the oscillating field can then be accounted for by using $B_\mathrm{tot}^2=B_0^2+\langle B^2\rangle$ in the assessment of quadratic Zeeman shifts.  This appears to be commonly used in the assessment of magnetic field effects in high accuracy clocks today \cite{AlThesis, AlIon1,PTbThesis,PeterBlythe}.  However it must be stressed that it only applies when the oscillating field is collinear with the static field.  

Coupling between fine-structure levels can be treated in exactly the same way as for the electric dipole polarisability and all expressions given in \cite{polarisability} have a magnetic analogue.  However, this is not the case within a fine-structure manifold.  Oscillating fields collinear with a static magnetic field couple only to neighbouring hyperfine states with $\Delta F=\pm 1, \Delta m=0$.  As hyperfine splittings are often much larger than frequencies of interest, the static limit applies and the static quadratic shift can be used as noted above.  Hence, results here concern the influence of ac magnetic fields orthogonal to the applied static field.

\subsection{Coupling within a single hyperfine level}
\label{sectTheoryHF}
The most straightforward case is the coupling between neighbouring $m$ states of the same hyperfine level.  From the Wigner-Eckart theorem
\begin{multline}
|\bra{F, m\pm1}\mathbf{e}_\pm\cdot \mathbf{M}\ket{F,m}|^2\\=\frac{(g_F \mu_B)^2}{2} (F\mp m)(F+1\pm m)
\end{multline}
and \eref{eq1} gives
\begin{equation}
\frac{\delta E}{\hbar}=\pm\frac{\langle B_\pm^2\rangle}{2} \frac{m}{\pm\omega_z-\omega}\left(\frac{g_F \mu_B}{\hbar}\right)^2,
\end{equation}
where $\omega_z=g_F \mu_B B_0/\hbar$ and $B_\pm$ are the spherical components of $B$.  Further assuming the field is linearly polarised, the contributions from each circular component are equally weighted giving
\begin{equation}
\label{ZeemanEff}
\frac{\delta E}{\hbar}=\left[\frac{1}{2} \frac{\omega_z^2}{\omega_z^2-\omega^2}\frac{\langle B_\perp^2\rangle}{B_0^2}\right] m \omega_z.
\end{equation}
This is a modification of the Zeeman shift $m \omega_z$, with the term in square parentheses having the interpretation of a fractional change in the applied magnetic field.  In the dc limit ($\omega\rightarrow 0$), the rms value is formally replaced by the amplitude and the shift is the modification of the field amplitude due to a static field applied in the transverse direction.  

Strictly speaking, Eq.~\ref{ZeemanEff} only applies for frequencies significantly different from the Zeeman splitting.  If this is not the case, population dynamics must be properly accounted for. On or near resonance there will be a Larmor precession of the spin. As demonstrated in section \ref{sect_AT}, this provides an accurate means to measure $B_\perp$. 

\subsection{Microwave clock transitions}
Consider an $S_\mathrm{1/2}$ ground-state with a half integer nuclear spin $I$.  Using the Wigner-Eckart theorem, the shift in energy $\hbar\delta\omega_\pm$ of $\ket{I\pm1/2,0}$ is
\begin{multline}
\hbar\delta\omega_\pm=\pm\frac{\omega_0}{\omega_0^2-\omega^2}\frac{|\bra{S_{1/2}}|\mathbf{M}|\ket{S_{1/2}}|^2}{\hbar}\\
\times\left(\frac{1}{6}\langle B_z^2\rangle+\frac{2I+1\mp2}{12(2I+1)}\langle B_\perp^2\rangle\right).
\end{multline}
As $|\bra{S_{1/2}}|\mathbf{M}|\ket{S_{1/2}}|^2\approx 6\mu_B^2$, the net shift of the clock transition is
\begin{equation}
\label{microwaveEq}
\delta\omega_0=\frac{|\omega_0|}{\omega_0^2-\omega^2}\frac{\mu_B^2}{\hbar^2}\left(2\langle B_z^2\rangle+\langle B_\perp^2\rangle\right).
\end{equation}
Note that the approximation for the reduced matrix element uses $g_J\approx2$ and neglects $g_I$.  The clock shift varies by a factor of 2 depending on the orientation of the oscillating field.  In the limit that $\omega \ll |\omega_{0}|$, the expression reduces to 
\begin{equation}
\delta\omega_0=\alpha_z \left(\langle B_z^2\rangle+\frac{1}{2}\langle B_\perp^2\rangle\right),
\end{equation}
where $\alpha_z$ is the quadratic shift of the clock transition due to a dc field.  

This is particularly relevant for the assessment of the magnetic field shifts in the Al$^+$ clock at NIST.  In their experiments, the oscillating field from rf currents induced in the trapping electrodes was measured by determining the shift of the microwave clock transition in either Be$^+$ or Mg$^+$ as the rf drive power is varied.  As discussed in \cite{AlThesis}, the analysis is based on the Breit-Rabi formula, which is equivalent to assuming the orientation is along $z$.  Consequently the inferred contribution could be two times larger.  From the numbers given in \cite{AlIon1} this would be an error of $1.4\times 10^{-18}$ in their clock assessment.  While this doesn't significantly change the total systematic uncertainties of the clocks reported in  \cite{AlIon1}, future Al$^+$ clocks with total uncertainty near $10^{-18}$ will need to take this into account.
\subsection{Optical clock transitions}
The analysis can be easily applied to other hyperfine structures.  As noted earlier, coupling between fine-structure levels can be treated as for an electric dipole polarizability \cite{polarisability}.  In the limit that the detuning is large relative to the hyperfine splitting of the upper state, the shift can be broken down into scalar, vector and tensor components.  The vector term only applies for circularly polarised field components and even then do not apply for $m=0$ states or cancel when averaged over Zeeman states with $m$ values of opposite sign.  The tensor term has a similar dependence as for an electric polarisability. In Lu$^+$ for example, coupling to the $^3D_2$ level gives a shift for each clock state in $^3D_1$ of
\begin{multline}
\label{finestructure}
\Delta \omega_F=-\left(\frac{1}{9}|\bra{{^3D_2}}|\mathbf{M}|\ket{{^3D_1}}|^2\right)\frac{\omega_\mathrm{fs}}{\omega_\mathrm{fs}^2-\omega^2}\frac{\langle B^2\rangle}{\hbar^2} \\
\times\left(1-\frac{C_{2,F}}{20}\left(3\cos^2\theta-1\right)\right),
\end{multline}
where $\theta$ is the angle between the ac field direction and the quantization axis, $\omega_\mathrm{fs}$ is the fine-structure splitting and $C_{2,F}$ is a coefficient that depends only on the angular momentum quantum numbers for the state of interest.  Under various averaging schemes \cite{ItanoQuad, dube2005electric, barrett2015NJP}, only the usual scalar term remains, which has the same quadratic dependence as for a static field in the limit that $\omega\ll \omega_\mathrm{fs}$. 

For clock transitions involving levels with a hyperfine structure, such as Yb$^+$, Hg$^+$, and Lu$^+$, the clock shift also has an orientation dependence not cancelled by averaging.  For Hg$^+$, the clock shift is given by
\begin{multline}
\delta \omega_c = \alpha_z(D_{5/2},2,0)\left(\langle B_z^2\rangle+\frac{2}{3}\langle B_\perp^2\rangle\right)\\
-\alpha_z(S_{1/2},0,0)\langle B^2\rangle,
\end{multline}
where $\alpha_z(D_{5/2},2,0)$ and $\alpha_z(S_{1/2},0,0)$ are the static quadratic Zeeman shift coefficient for the upper and lower clock states, respectively.  Note that the shift for the lower state is proportional to $\langle B^2\rangle$, which is a consequence of its zero angular momentum.  The clock frequency is averaged over three orthogonal field directions, which replaces each component with one third of the total, giving
\begin{equation}
\delta \omega_c = \left(\frac{7}{9}\alpha_z(D_{5/2},2,0)-\alpha_z(S_{1/2},0,0)\right)\langle B^2\rangle.
\end{equation}
This gives $\sim -13.7\,\mathrm{mHz/\mu T^2}$ compared to the static value of $\sim -19.0\,\mathrm{mHz/\mu T^2}$ calculated in \cite{ItanoQuad}.  The averaging therefore restores the assumed dependence on $\langle B^2\rangle$ albeit at a modified shift coefficient.  This would not affect the order of magnitude estimate given in \cite{AlHgComparison}.

A similar consideration applies to Yb$^+$.  However, owing to a near cancellation of the quadratic Zeeman coefficients for the upper and lower states, the effect is more pronounced.  The clock shift after averaging is given by
\begin{equation}
\delta \omega_c = \left(\frac{3}{4}\alpha_z(F_{7/2},3,0)-\alpha_z(S_{1/2},0,0)\right)\langle B^2\rangle,
\end{equation}
From the values of hyperfine splittings given in \cite{PeterBlythe}, the coefficient is $2.24\,\mathrm{mHz/\mu T^2}$ compared to $-2.18\,\mathrm{mHz/\mu T^2}$ for the static case.  Thus the correction effectively has the wrong sign when simply adding $\langle B^2\rangle$ as suggested in \cite{PeterBlythe}.  It is unclear how much this would affect clock assessments as reports \cite{YbIon1, YbIon2, YbIon3, PeterBlythe} do not elaborate on how or if the ac fields are assessed.  Measured quadratic Zeeman coefficients vary substantially with values differing by as much as $12\sigma$ of the claimed uncertainties \cite{YbIon2,YbIon3}, but it is not always stated what value is being used.  The most current and accurate value of the quadratic shift coefficient is given in \cite{YbIon3}, but the reported clock shifts are consistent with zero contribution from ac fields.  Although it may well be the case that rf currents are significantly reduced at different operating conditions, the sensitivity to ac currents is 30-fold larger for Yb$^+$ compared to Al$^+$. Thus it would seem prudent to consider this effect,  particularly in light of experiments investigating the variation of fundamental constants \cite{YbIon3,huntemann2014improved}. 

For lutetium, calculations can be easily extended to include more hyperfine levels.  For each level, the shift can be written
\begin{equation}
\Delta f_F=\alpha_F \langle B_z^2 \rangle+\alpha'_F \langle B_\mathrm{\perp}^2 \rangle.
\end{equation}
Under hyperfine averaging \cite{barrett2015NJP}, $\alpha_F$ averages to zero but not $\alpha'_F$.  In table \ref{LuShifts}, $\alpha$ and $\alpha'$ are listed for each hyperfine level of each clock transition and the hyperfine averaged $\alpha'$ is also given.  The values quoted are determined from measured hyperfine splittings and do not include the much smaller contributions from neighbouring fine-structure levels.  For comparison, the coefficients for other ion-based clocks, under the appropriate averaging schemes, are given in table \ref{SummaryTable}.  Clearly those candidates having a hyperfine structure are significantly more sensitive in general and the value for the 848-nm transition in $^{176}$Lu$^+$ may seem anomalously small in this regard.  This is owing to a fortuitous hyperfine structure that balances the splittings and suppresses the shift.

\begin{table}[h]
 \caption{Quadratic Zeeman shift coefficients for ac magnetic fields for clock transitions in $^{176}$Lu$^+$: $\alpha_F$ applies to fields aligned along the quantisation axis, $\alpha_F'$ applies to perpendicular fields.  All values are expressed in $\mathrm{mHz/\mu T^2}$.}
 \label{LuShifts}
\centering
\begin{tabular}{c c c c}
\toprule
 & $F$ & $\alpha_F (\mathrm{mHz/\mu T^2}) $ & $\alpha'_F (\mathrm{mHz/\mu T^2})$\\
 \hline
 $^3D_1$ & 6 & 2.32 & 1.32\\
 & 7 & -0.14 & 0.23\\
 & 8 & -2.18 & -0.95\\
 & $\ev{\cdot}_F $ & - & 0.20\\
 \hline
$^3D_2$ & 5 & -44.03 & -25.68\\
 & 6 & -1.71 & -7.79\\
 & 7 & 12.54 & 0.93\\
  & 8 & 16.57 & 5.29\\
 & 9 & 16.62 & 7.39\\
 & $\ev{\cdot}_F $ & - & -3.98\\
 \hline
$^1D_2$ & 5 & 53.45 &31.18\\
 & 6 & 14.61 & 16.62\\
 & 7 & -3.53 & 7.13\\
  & 8 & -17.73 & -2.23\\
 & 9 & -46.80 & -20.80\\
 & $\ev{\cdot}_F $ & - & 6.38\\
 \hline
 \end{tabular}
 \end{table}
 
 \begin{table}[!ht]
\caption{The quadratic ac magnetic field sensitivities and fractional shifts of different optical frequency standards.  For $^{176}$Lu$^+$ the dependence is on $\langle B_\perp^2 \rangle$.  All others depend on $\langle B^2 \rangle$.}
\label{SummaryTable}
\begin{ruledtabular}
\begin{tabular}{cccc}
Ion & $\lambda$ (nm) & $\tilde{\alpha}_z$ ($\mathrm{mHz/\mu T}^2$) & $\delta f/f$ ($\mathrm{\mu T}^{-2}$) \\
\hline
$^{199}$Hg$^+$  	& 282 & -13.7 \footnote{Averaged over three orthogonal axes \cite{ItanoQuad}. \label{Bav}} & $-1.3\times 10^{-17}$ \\
$^{171}$Yb$^+$ E2  & 436 & 33.8 \textsuperscript{\ref{Bav}} & $4.9\times 10^{-17}$ \\
$^{171}$Yb$^+$ E3  & 467 & 2.28 \textsuperscript{\ref{Bav}} & $3.5\times 10^{-18}$ \\
$^{88}$Sr$^+$  	& 674 & 0.0031 \footnote{Averaged over Zeeman states \cite{dube2005electric}. \label{Mav}} & $7.0\times 10^{-21}$ \\
$^{40}$Ca$^+$  	& 729 & 0.014 \textsuperscript{\ref{Mav}} & $3.5\times 10^{-20}$ \\
$^{27}$Al$^+$  		& 267 & -0.072 & $-6.4\times 10^{-20}$ \\
$^{115}$In$^+$  	& 236 & -0.004 & $-3.2\times 10^{-21}$ \\
$^{176}$Lu$^+$ $({^3}D_1)$ & 848 & 0.20 \footnote{Hyperfine averaging \cite{barrett2015NJP}.  For these transitions, dependence is on $\langle B^2_\perp\rangle$ \label{HFav}} & $5.7\times 10^{-19}$ \\
$^{176}$Lu$^+$ $({^3}D_2)$ & 804 & -3.98 \textsuperscript{\ref{HFav}} & $-1.1\times 10^{-17}$ \\
$^{176}$Lu$^+$ $({^1}D_2)$ & 577 & 6.38 \textsuperscript{\ref{HFav}} & $1.2\times 10^{-17}$ \\
\end{tabular}
\end{ruledtabular}
\end{table}

\subsection{Blackbody magnetic fields}
\begin{figure}[h]
\begin{center}
  \includegraphics[width=3in]{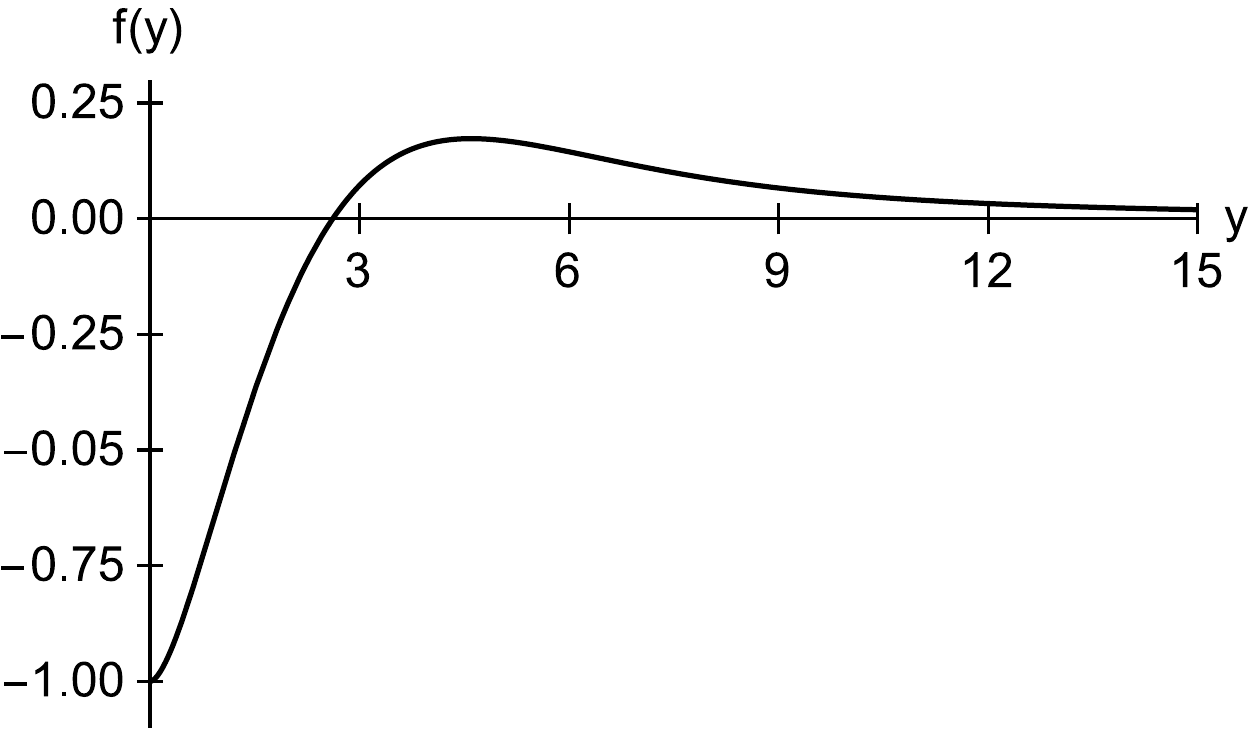}
  \caption{The figure shows the the function $f(y)$ where $y=\hbar \omega_0/k_B T$.}
  \label{plot_f}
\end{center}
\end{figure}

Blackbody radiation also provides a shift contribution from the thermal magnetic field.  For optical transitions this is much less significant than the shift from thermal electric fields but we include it here for completeness.  The thermal magnetic field has a mean squared value of
\begin{align}
\langle B^2(t) \rangle&=\frac{\hbar}{\pi^2 c^5\epsilon_0} \int_0^\infty\frac{\omega^3 d\omega}{\exp\left(\frac{\hbar \omega}{k_B T}\right)-1}\\
&\approx \left(2.77507\,\mathrm{\mu T}\right)^2\left(\frac{T}{T_0}\right)^4,
\end{align}
where $T_0=300\,\mathrm{K}$.  For a given transition it is useful to note that
\begin{align}
\frac{\delta \omega_0}{\omega_0}&=\frac{\mu_B^2}{\hbar\pi^2 c^5 \epsilon_0} \int_0^\infty \frac{1}{\omega_{0}^2-\omega^2} \frac{\omega^3 d\omega}{\exp\left(\frac{\hbar \omega}{k_B T}\right)-1}\\
&=\frac{\mu_B^2}{\hbar\pi^2 c^5 \epsilon_0} \left(\frac{k_B T}{\hbar}\right)^2 \int_0^\infty \frac{1}{y^2-x^2} \frac{x^3 dx}{e^x-1}\\
&= -\beta \left(\frac{T}{T_0}\right)^2f(y),
\label{magshift}
\end{align}
where $y=\hbar \omega_0/(k_B T)$,
\begin{equation}
\beta=\frac{\mu_B^2}{\hbar^2} \frac{\hbar}{6 c^5 \epsilon_0} \left(\frac{k_B T_0}{\hbar}\right)^2\approx 9.78\times 10^{-18},
\end{equation}
and
\begin{equation}
f(y)=\frac{6}{\pi^2}\int_0^\infty \frac{1}{y^2-x^2} \frac{x^3 dx}{e^x-1}.
\end{equation}
The integral is to be interpreted as the principle value and is plotted in Fig.~\ref{plot_f}.  

For $S_{1/2}$ microwave clock transitions the fractional shift is
\begin{equation}
\frac{\delta \omega_0}{\omega_0}\approx 1.304\times 10^{-17}  \left(\frac{T}{T_0}\right)^2,
\end{equation}
where we have used the fact that $f(y)\approx f(0)=-1$ and the radiation field is isotropic.  This result is in agreement with \cite{ItanoMagneticBBR} and explicitly relies on the validity of Eq.~\ref{microwaveEq} 

In the expressions above $\omega_0$ is the transition frequency of the contributing M1 transition.  For an optical clock transition, the fractional frequency shift is suppressed by a further factor of $\omega_0/\omega_c$. Thus shifts from coupling between hyperfine levels is negligible and we need only consider coupling to other fine-structure levels.  Even in this case, fine-structure splittings are typically one to two orders of magnitude smaller than the optical transition, and there is a further suppression due to $f(y)$ for the larger splittings.  Hence magnetic BBR shifts are not likely to be significant in any realistic scenario.  

To illustrate, the $^{176}$Lu$^+$ fine-structure splitting between $^3D_1$ to $^3D_2$ is approximately 19.2\,THz giving, $y\approx 3.06$ at $T=300\,\mathrm{K}$ and $f(y)\approx0.1348$.  The corresponding shift of the 848-nm optical clock transition is then $-3.35\times 10^{-20}$ or $-3.48\times 10^{-20}$ when including the contribution from $^1D_2$.  Note that $y$ itself is a function of temperature so this shift is not simply quadratic in temperature as indicated by Eq.~\ref{magshift}.  Shift of the 804-nm clock transition is similarly found to be $-3.22\times 10^{-20}$ which includes coupling to all other $D$-states.

\section{Measuring ac magnetic field shifts}
With the ever-increasing precision of optical clocks and measurements carried out in ion-trap systems, it would be ideal to have a technique to precisely measure the amplitude and orientation of various oscillating fields, specifically the trap-induced rf fields.  A standard technique has been to vary the rf confinement and extrapolate any measurable difference to zero \cite{berkeland2002laser,AlIon1, AlThesis, hoffman2013radio, lewty2013experimental}.  This is not always ideal as averaging times can be very long and a more direct approach would be better.  In this section we discuss two complementary approaches: one based on an Autler-Townes splitting induced by $B_\perp$ \cite{autler1955stark}, and the other on a sideband induced by $B_z$ \cite{meir2018experimental}.

\subsection{Autler-Townes splitting from an ac magnetic field.}
\label{sect_AT}
As noted in section~\ref{sectTheoryHF}, matching the Zeeman splitting to the trap drive rf can result in a Larmor precession.  When driven on a connected optical or microwave transition, an Autler-Townes splitting arises \cite{autler1955stark}.  The splitting can be measured accurately and is a direct measure of $B_\perp$.  This approach is readily applicable when there is an available energy level with an appreciable $g$-factor and a moderate trap drive frequency.  Here we demonstrate this technique using $^{171}\mathrm{Yb}^+$ confined in a linear Paul trap.  In this system, the $F=1$ ground-state hyperfine level has $g_F\approx1$, and the Zeeman splitting can be matched to the trap drive frequency of $\Omega_\mathrm{rf}=2\pi\times30.1891\,\mathrm{MHz}$ with a readily achievable field of $\sim2.15\,\mathrm{mT}$.

\begin{figure}
\begin{center}
	\includegraphics[width=3.25in]{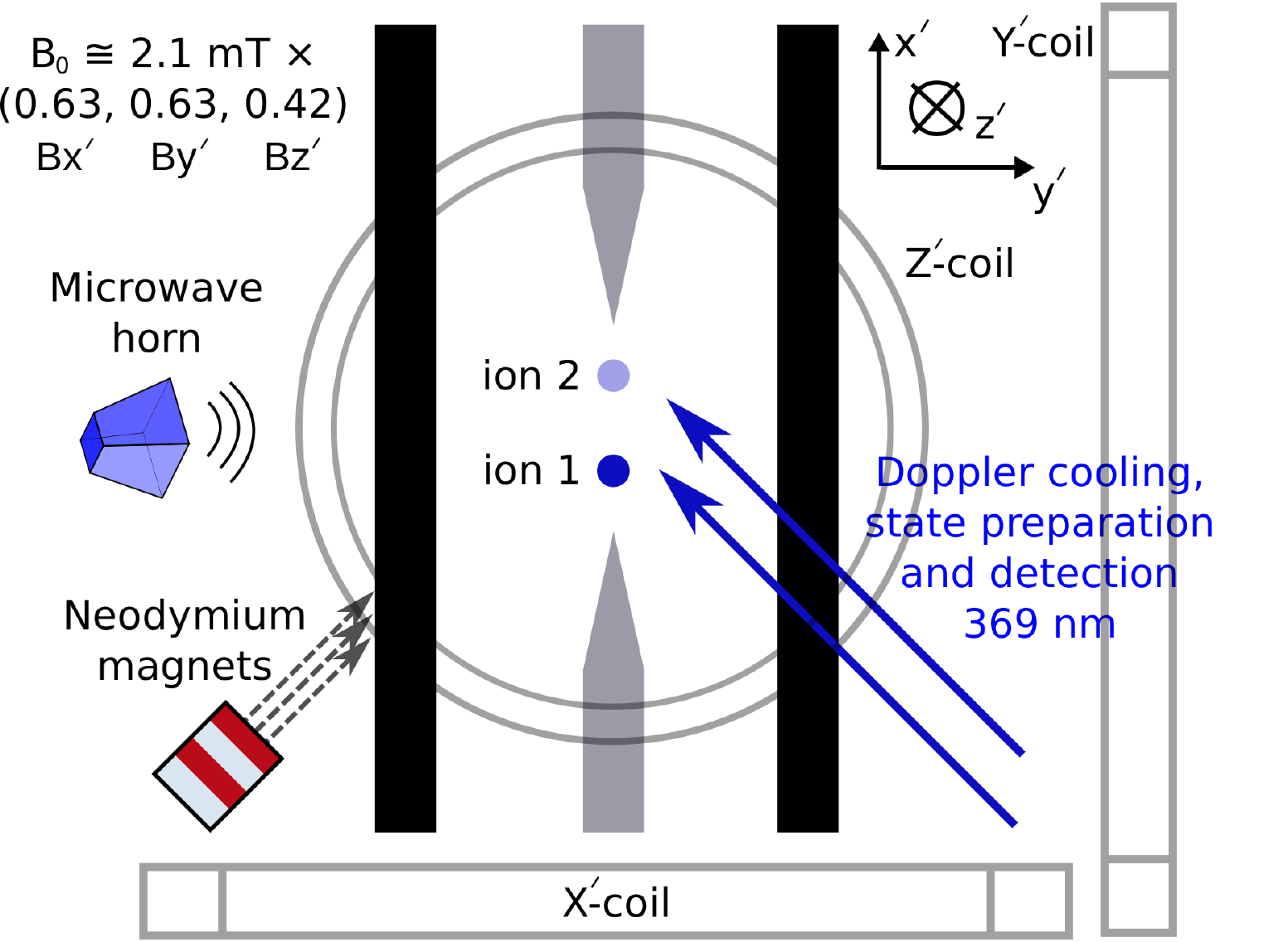}
	\vspace{0.25cm}\\
  \includegraphics[width=3.25in]{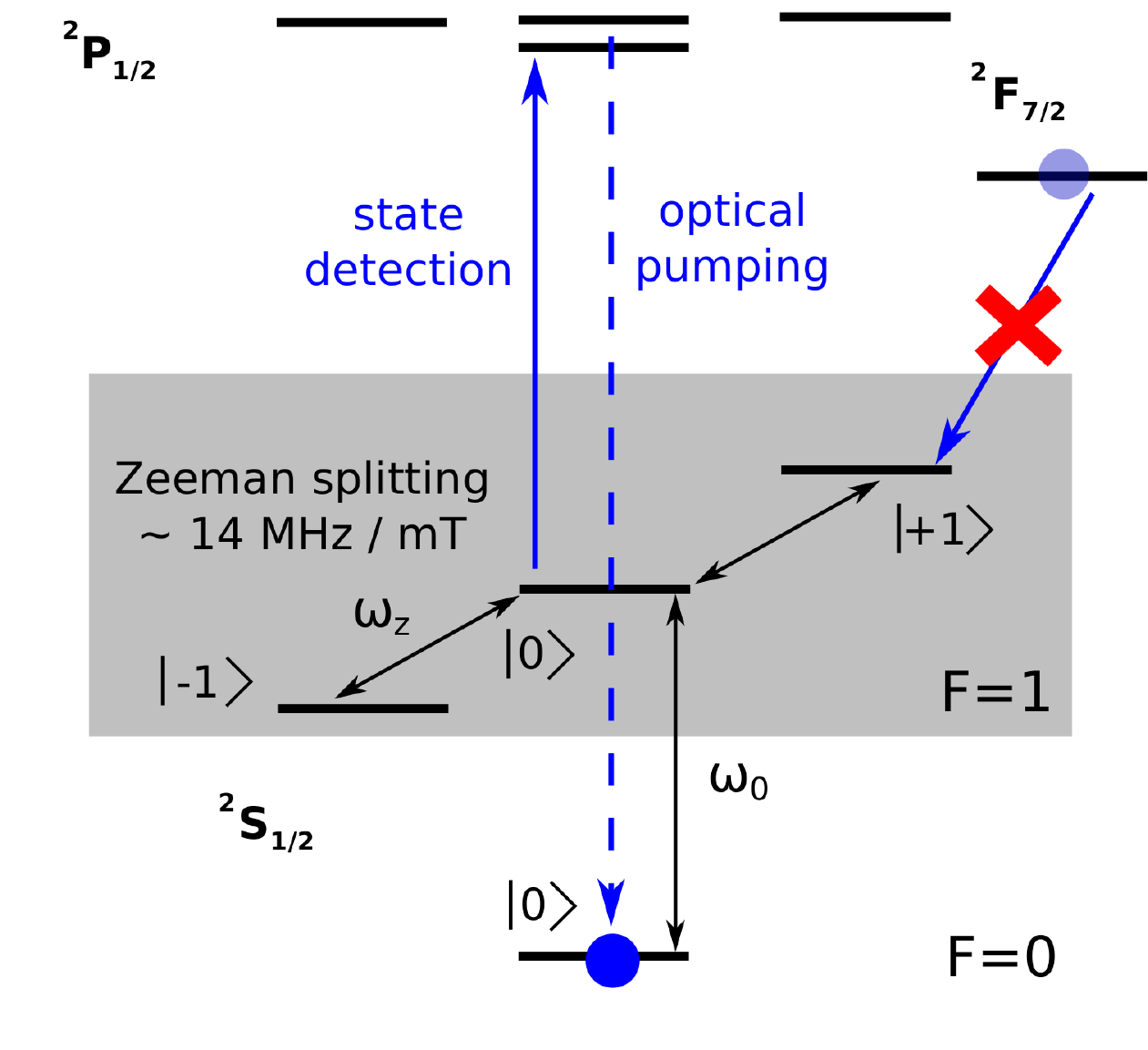}
  \caption{Schematic of the experimental setup.  A small stack of neodymium magnets is used to augment an existing field to create a bias field of $B_0\approx 2.1\,\mathrm{mT}$ along the direction (0.63, 0.63, 0.42).  The $Z^{\prime}$coil current $i_z$ is used to fine tune $B_0$ with $< 1^\circ$ change in the field direction over the small scan range of interest. A microwave horn is used to drive microwave transitions between the $F=0$ and $F=1$ levels, which has a zero-field separation of $\omega_0 = 12,642,812,118\,\mathrm{Hz}$~\cite{Fisk_Yb_hyperfine_measurement_1997}. The first-order Zeeman effect gives $\approx$ 14 MHz/mT for the frequency shift of $\ket{1, \pm 1}$ with respect to $\ket{1, 0}$. The quadratic shift of the $\ket{0,0}\rightarrow\ket{1,0}$ transition is $\alpha_z=31080\,\mathrm{Hz/mT^2}$~\cite{Vanier_Quantum_Physics_of_Atomic_Freq_Standards_1989}.
}
  \label{YbExpt}
\end{center}
\end{figure}

The experiment is carried out in a four rod linear Paul trap with axial end caps as described in \cite{ding2014microwave,ding2017quantum}. 
The trap geometry and relevant level structure are schematically shown in Fig~\ref{YbExpt}. The secular trap frequencies for a single ion are $(\omega_{r1},\omega_{r2},\omega_{ax})/2\pi = (0.539,0.857,0.251)\,\text{MHz}$ for two radial and axial trapping directions respectively.  Doppler cooling, detection, and state preparation are carried out via scattering to the $P_{1/2}$ level as described in \cite{olmschenk2007manipulation}.  Microwave transitions between the $F=0$ and $F=1$ levels are driven using a microwave horn located $\sim 5\,\mathrm{cm}$ from the trap center.

A small stack of neodymium magnets placed approximately 13\,cm from the trap center is used to augment an existing field of approximately $0.6\,\mathrm{mT}$ so that the $F=1$ Zeeman splittings near match the trap drive frequency.  The combined field of $B_0 \approx 2.1\,\mathrm{mT}$ has a direction vector $\sim (0.63, 0.63, 0.42)$ with respect to the coordinate system shown in Fig~\ref{YbExpt}. The axes are primed to avoid possible confusion with notation introduced earlier for the ac-field components.  The $Z^{\prime}$-coil current ($i_z$) is used to fine tune the amplitude of the magnetic field.  Over the small tuning range used, this primarily changes the amplitude of $B_0$ by approximately $-0.094\,\mathrm{\mu T/mA}$ with only a small change of approximately $\pm 0.6^\circ$ in the direction of the field.

The experimental sequence is as follows: for each value of $i_z$, the ion is first Doppler cooled and optically pumped into the $\ket{0,0}$ hyperfine ground state.  A 100 $\mu$s microwave pulse is then used to drive the atom to the $F=1$ level.  Successful transfer to $F=1$ is determined from fluorescence collected during resonant excitation of the $S_{1/2}$, $F=1$ to $P_{1/2}$, $F=0$ transition and the transfer probability is inferred from 100 experiments. The amplitude of the microwave drive is chosen to maximize the resonant population transfer for the target $m$ state of interest.   

\begin{figure}
\begin{center}
  \includegraphics[width=3.0in]{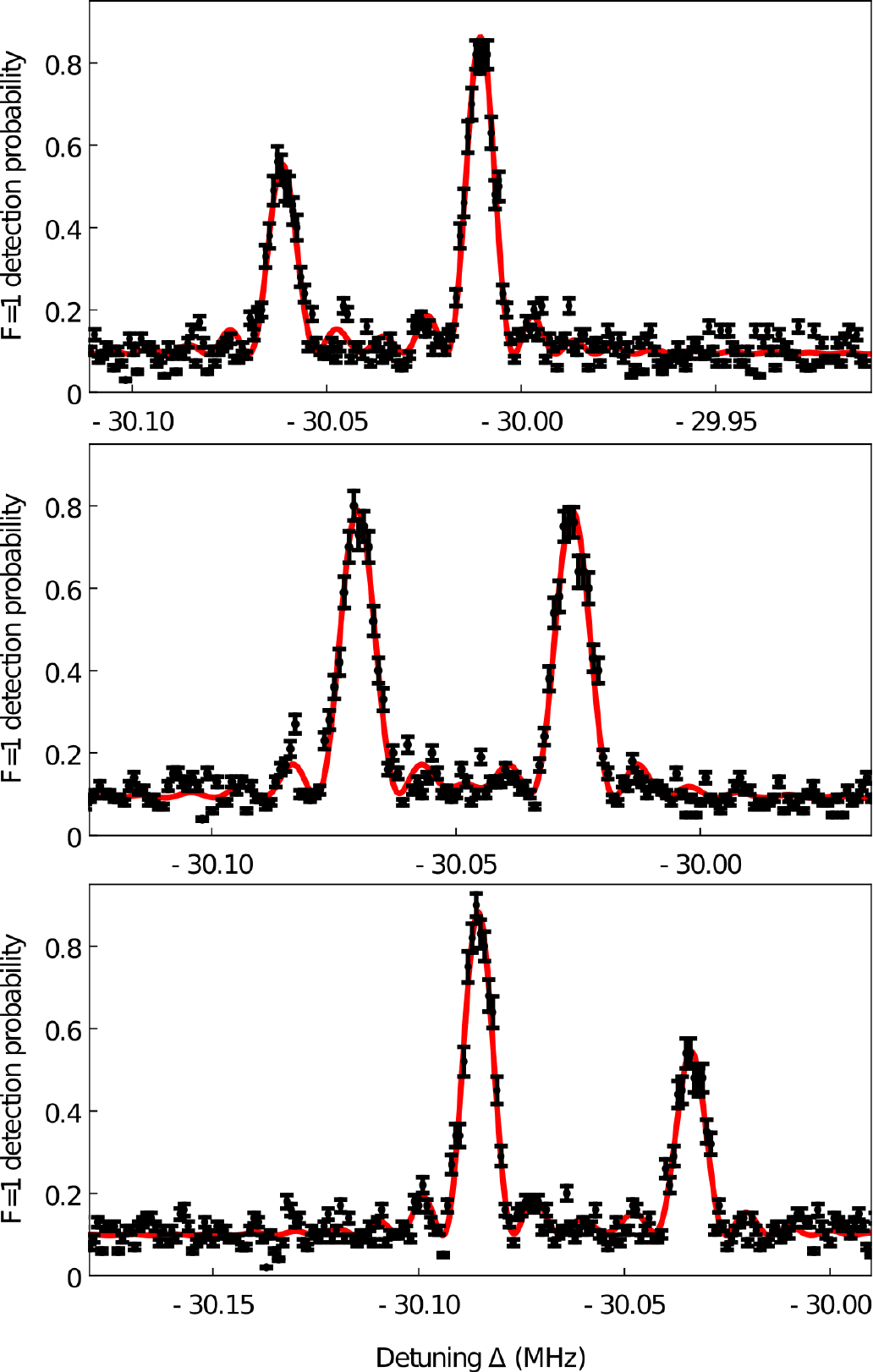}
 \caption{Microwave frequency scans at fixed $i_z$ ($B_0$).  Detunings are given relative to the zero field ground-state hyperfine splitting.   A trap-induced magnetic coupling results in an Autler-Townes splitting when the Zeeman splitting, $\omega_z$, between $\ket{1,-1}$ and $\ket{1,0}$ is near to $\Omega_\mathrm{rf}$.  The two peaks correspond to the two dressed states arising from the coupling. Top, middle, and bottom traces correspond to $\omega_z<\Omega_\mathrm{rf},\,\omega_z \approx \Omega_\mathrm{rf},$ and $\omega_z>\Omega_\mathrm{rf}$, respectively.  Note the slight difference in the horizontal axis in each case.}
  \label{scan}
\end{center}
\end{figure}

Typical microwave frequency scans for fixed $i_z$ are shown in Fig.~\ref{scan}.  When the Zeeman splitting, $\omega_z$, between $\ket{1,-1}$ and $\ket{1,0}$ is near to $\Omega_\mathrm{rf}$, an Autler-Townes splitting occurs with the two peaks corresponding to the two dressed states~\cite{Cohen-Tannoudji1996} arising from the trap-induced magnetic coupling. For $\omega_z=\Omega_\mathrm{rf}$ the peaks are symmetric and the splitting is determined by the strength of the coupling.   As $\omega_z$ is tuned away from $\Omega_\mathrm{rf}$, the peaks become asymmetric with a larger separation, and the dominant peak moves towards the energy of the bare eigenstate.  

\begin{figure}
\begin{center}
  \includegraphics[width=0.45\textwidth]{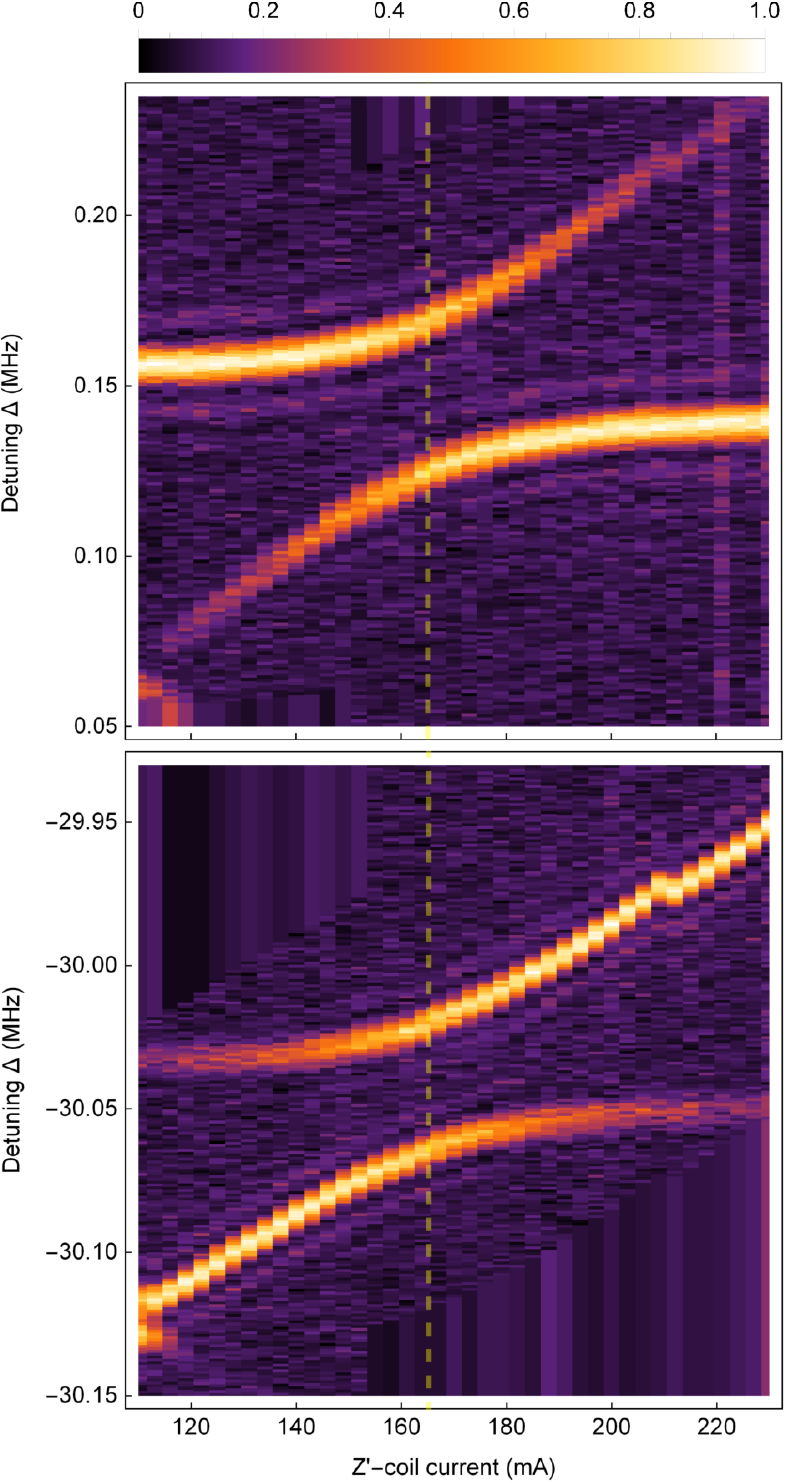}
 \caption{ Observed Autler-Townes splittings as a function of $i_z$ for microwave frequencies near to the bare resonances associated with $\ket{1,0}$ and $\ket{1,-1}$  for the upper and lower plots, respectively. Microwave frequency on the vertical axis is relative to the $^{171}\mathrm{Yb}^+$ zero field ground-state hyperfine splitting. The vertical dashed line corresponds to the value of $i_z$ at the minimum splitting as determined from the data: $2\pi\times 44.8(3)\,\mathrm{kHz}$ in both cases.}
  \label{fullscan}
\end{center}
\end{figure}

Scans over a range of magnetic fields are shown in Fig.~\ref{fullscan} for microwave frequencies near to the bare resonances associated with $\ket{1,0}$ and $\ket{1,-1}$  for the upper and lower plots, respectively. A small splitting seen at the bottom left of the figure is due to a two-photon coupling between $\ket{1,\pm1}$ when the Zeeman splitting between these two states is $\sim2\Omega_\mathrm{rf}$.  Over the magnetic field range used, no other splitting near to the $\ket{1,1}$ bare resonance is observed due to a $\sim 72\,\mathrm{kHz}$ quadratic Zeeman shift of $\ket{1,0}$.  

Neglecting the contribution from $\ket{1,1}$, the observed splitting is given by $\sqrt{\delta^2 + \Omega^2}$ where $\delta=\omega_z-\Omega_{\mathrm{rf}}$ and $\Omega$ is the coupling strength between $\ket{1,-1}$ and $\ket{1,0}$ due to the trap-induced rf magnetic field.  Using $\Omega$ as a fit parameter, gives $\Omega = 2\pi\times 44.8(3)\,\mathrm{kHz}$ for both sets of data in Fig.~\ref{fullscan}. 
Assuming the rf magnetic field is linearly polarized, the coupling strength is given by
\begin{equation}
\label{coupling}
\Omega=\frac{g_F \mu_B B_\perp}{\hbar\sqrt{2}}.
\end{equation}
Using $g_F=1$ and the measured splitting then gives $B_\perp=4.527(30)\,\mathrm{\mu T}$.  

The measured splitting is largely independent of the calibration of $B_0$.  Moreover it is also insensitive to the exact values of $g_F$ and $\alpha_z$.  These parameters determine the location of the splitting but the size of the splitting is almost entirely determined by $\Omega$.  The small dependence on $g_F$ and $\alpha_z$ comes from the location and strength of the nearby one- and two-photon resonances associated with $\ket{1,1}$.  Inclusion of this state in the analysis shifts the estimated coupling to $\Omega=2\pi\times45.3(3)\,\mathrm{kHz}$ with a corresponding change to $B_\perp$ in accordance with Eq.~\ref{coupling}.  Further corrections due to errors in $g_F$, and $\alpha_z$ are less than the error in determining the minimum value.

Strictly speaking the splitting only depends on the $\mathbf{e}_+$ component of the magnetic field.  In principle the $\mathbf{e}_-$ component could be checked by reversing the field.  However an imbalance in the weight of each component would imply a significant phase shift between contributing current sources that would likely be associated with substantial micromotion that could not be compensated by bias fields.

\begin{figure}
\begin{center}
  \includegraphics[width=0.45\textwidth]{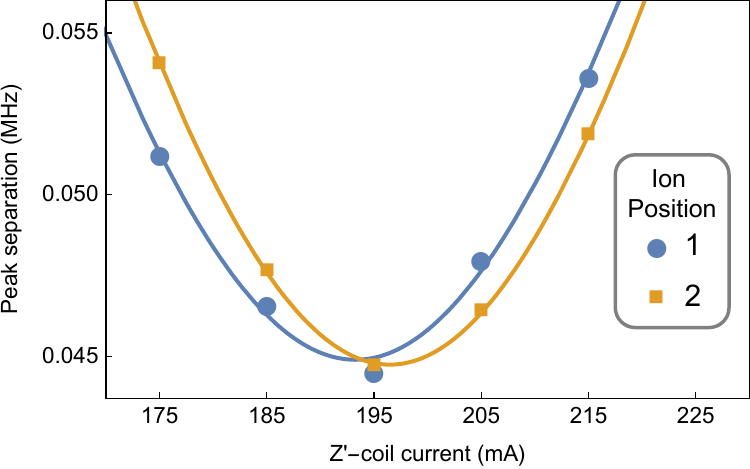}
 \caption{Measured Autler-Townes splitting as a function of $i_z$ with two ions in the trap. In each case one ion is shelved to the $^{2}F_{7/2}$ dark state throughout the scan. The bright ion was always kept at position 1 (blue dots) or position 2 (orange squares) as shown in Fig.~\ref{YbExpt}. The curves are the fits to the two level result as discussed in the text.  Displacement of the plots is due to a spatial inhomogeneity in $B_0$.  However the minimum splitting and hence $B_\perp$ is fairly constant.}
  \label{TwoIons}
\end{center}
\end{figure}

As the rf-currents are driven by the trapping fields themselves, a spatial dependence to the ac magnetic field can be expected.  To investigate this, a second ion in the long lived $^{2}F_{7/2}$ level was used to displace the first along the trap axis with the separation between ions estimated to be $8.7\,\mu$m. The measured splitting as a function of magnetic field for the bright ion at either position along the trap axis is shown in Fig.~\ref{TwoIons}.  The displacement of the two plots indicates that $B_0$ has a gradient along the trap axis of about 35\,mT/m but $B_\perp$ remains fairly constant.

\begin{figure}
\begin{center}
  \includegraphics[width=0.45\textwidth]{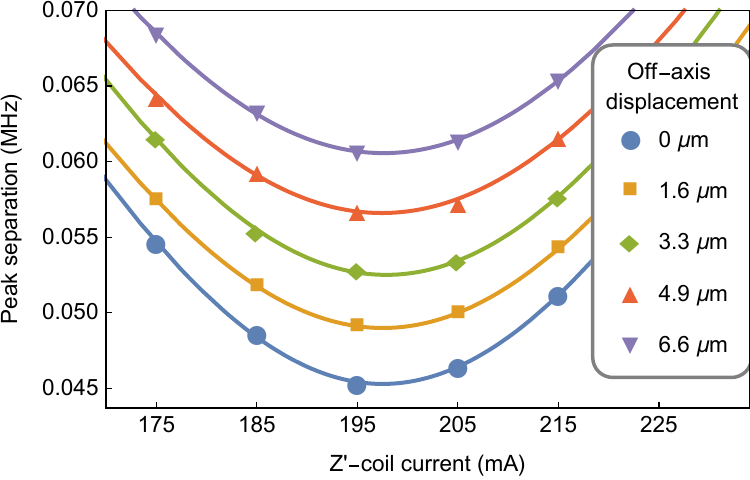}
 \caption{Measured Autler-Townes splitting as a function of $i_z$ as the ion is moved off-axis in the radial direction.  Distances are calibrated from camera images with an accuracy of $\sim20\%$.  As expected, there is a significant change in the splitting and hence $B_\perp$.}
  \label{micromotion}
\end{center}
\end{figure}

A more significant variation in $B_\perp$ can be expected for displacements off-axis. This was investigated by applying a dc bias voltage to one of the rf trap electrodes to move the ion off-axis. The results are shown in Fig~\ref{micromotion}, in which displacements were inferred from camera images with an accuracy of $\sim20\%$.  From the data, $\Omega$ has an approximately linear dependence of $\sim2.3\,\mathrm{kHz/\mu m}$ on the ion displacement.  The linear dependence is expected from the four rod geometry of the trap, and suggests a zero in the ac magnetic field $\sim20\,\mathrm{\mu m}$ from the trap center.  This is not unreasonable given the machining and fabrication tolerances involved in trap construction.  


In the fortuitous event that the Autler-Townes splitting is too small to be resolvable, Larmor precession would then apply.  Two $\pi$-pulses on the $\ket{0,0}$ to $\ket{1,0}$ microwave transition separated by a time $\tau$, would see an oscillation of population in $F=1$ as a function of $\tau$ with a timescale determined by $\Omega$.  In general, the resulting signal might be more complicated depending on how much the quadratic Zeeman shift splits the degeneracy of the two Zeeman splittings.  For $J=0$ to $J=0$ transitions this would allow one to quantify line noise if the splittings could be tuned to near the line noise frequency and/or its first harmonic as done in \cite{boyd2007nuclear}.

\subsection{Magnetic field induced sidebands}
An alternative approach for measuring the ac field is to utilize a magnetic field-induced sideband.  First observed in \cite{meir2018experimental}, this effect can bias micromotion compensation as it also contributes to the sideband signal.  Alternatively, if micromotion is properly compensated, the residual sideband could then be attributed to the ac magnetic field. In contrast to the previous section, the effect depends on $B_z$.  

Far from resonance with a Zeeman splitting, $B_\perp$ effectively modifies the static field, whereas $B_z$ modulates the energy levels.  This modulation is formally equivalent to a phase modulation of the driving field with a modulation index given by
\begin{equation}
\label{MI}
\beta_m=\frac{(g_F' m_F'-g_F m_F)\mu_B B_z}{\hbar \omega}
\end{equation}
where the prime denotes excited state quantities and we have neglected any quadratic shifts.  Hence a measurement of the sideband to carrier ratio should allow $B_z$ to be extracted.  For this to be effective, other sources responsible for a signal at the sideband frequency must be eliminated or at the very least measured.  In the case of the rf sideband in ion-traps, this is predominantly micromotion, which has two components: excess micromtion (EMM) and intrinsic micromotion (IMM) \cite{berkeland1998minimization, keller2015precise}.

To disentangle the contribution from micromotion, it must be assessed and removed as much as possible.  The obvious strategy would be to first use a transition insensitive to magnetic fields to quantify the micromotion, and then use an alternative transition with a large $\beta_m$ to assess the magnetic field contribution.  Such a separation is not always possible as in the case of Sr$^+$.  In that case the techniques demonstrated in \cite{meir2018experimental} can be used.  Here we consider $^{176}$Lu$^+$ \cite{paez2016atomic, kaewuam2017laser} to illustrate the general considerations.

For $^{176}$Lu$^+$, the $^1S_0$-to-$^1D_2$ clock transition at 577\,nm is well-suited to micromotion assessment: power requirements for driving weak sidebands are reasonable, probing times of a few tens of ms are possible without significant decoherence, and the wavelength provides reasonable coupling to the motion.  Any of the clock transitions connected to an upper $m=0$ state, has a magnetic field sensitivity on the order of a few $\mathrm{Hz/\mathrm{\mu T}}$.  Moreover, two of the transitions are field independent at $\sim 0.1\,\mathrm{mT}$ with a quadratic dependence of $\sim15 \,\mathrm{mHz/\mu T^2}$.  At a trap drive frequency of $\sim 30\,\mathrm{MHz}$, $\beta_m$ is completely negligible for these transitions and the sideband signal limited only by micromotion.  However the $\ket{7,7}$ to $\ket{9,9}$ transition, which has the largest available magnetic sensitivity, has only a modest sensitivity of $\beta_m\sim 10^{-3}/\mathrm{\mu T}$.  This needs to be compared to the expected levels of micromotion compensation and how well the sidebands could be resolved.

A detailed account of micromotion limitations is given in \cite{keller2015precise}.  The minimum resolvable modulation index is limited by available laser power, laser coherence and IMM.  Probing along the trap axis of a linear Paul trap effectively eliminates IMM as the rf field amplitude along this direction is typically very small.  Coherence times on the $^1D_2$ clock transitions would only be limited by the upper state lifetime of $\sim 200\,\mathrm{ms}$ or thermal dephasing, which can be easily characterized.   With a laser power of $0.4\,\mathrm{mW}$ focussed to $30\,\mathrm{\mu m}$, a $25\,\mathrm{ms}$ probe resulting in a near 100\% transfer to the excited state at the rf sideband would correspond to a modulation index of $\sim10^{-3}$ for either transition.  So the accuracy at which assessment could be carried out would likely be determined by how well a $\pi$-time can be measured.  This is not likely to be as accurate as the determination of an Autler-Townes splitting.

In general, the accuracy via this technique is determined by the available $\beta_m$: the larger the better.  This implies a high $g$-factor and/or low trap drive frequencies both of which facilitate the achievement of magnetic fields necessary to observe an Autler-Townes splitting.  Nevertheless this approach may still be useful for those clocks in which the ac magnetic field shift is small.

\section{Discussion}
In this work the influences of ac magnetic fields have been explored and these should be carefully considered in any precision measurement.  For Paul traps, the trap-induced ac fields can be significant and should be considered a mandatory part of any realistic error budget.  For ion-based clocks, not only do the ac fields induce a shift in the clock frequency, they can also influence the proper assessment of micromotion as noted in \cite{meir2018experimental}.  This could further influence clock assessments if induced-micromotion is used to calibrate other systematics, for example, the blackbody radiation shift via the static differential polarizability as done in \cite{dube2014high}.  
 
More generally, magnetic field calibrations are often carried out by measuring Zeeman splittings.  Although the effect on Zeeman splittings is typically small, it can still be important in precision measurements.  A notable example is the high accuracy measurement of the $D_{5/2}$ $g_J$ factor in $^{40}$Ca$^+$, which was reported with a fractional inaccuracy of $2.5\times 10^{-7}$ \cite{chwalla2009absolute}.  This measurement relies on a comparison of Zeeman splittings between the $S_{1/2}$ and $D_{5/2}$ states.  In principle the ratio of the Zeeman splittings depends only on the ratio of $g$-factors between the two levels but Eq.~\ref{ZeemanEff} modifies that ratio.  The trap drive frequency was not given in the report, but a value of $20\,\mathrm{MHz}$ would give a sensitivity of $-8.3\times10^{-7}/\,\mathrm{\mu T^2}$ at the static magnetic field used in the experiment.

Possible methods to measure the ac field in an ion trap system have been discussed.  A simple approach using an observed Autler-Townes splitting demonstrated a $<1\%$ inaccuracy in the determination of $B_\perp$.  This method can be directly applied to Yb$^+$ clock experiments for which ac fields could be metrologically significant.  If not properly assessed in this system, it would also have significant repercussions for experiments testing the variation of fundamental constants \cite{YbIon3,huntemann2014improved}.  The method is also applicable to Hg$^+$ clock experiments, in particular the microwave clock \cite{berkeland2002laser}, for which the ac magnetic field shift was the leading systematic uncertainty.

For systems that require much larger fields to observe an Autler-Townes splitting, it may be possible to use a different species to first characterise the trap.  However this would depend on how stable and reproducible the effects are, and how they vary spatially.  Measurements shown here indicate the expected strong correlation with micromotion but this would have to be more extensively investigated in any given set up.  The alternative approach of using a magnetically induced sideband could also be used provided it could achieve sufficient accuracy.

As the ac currents in an ion trap are driven by the same source that determines the trapping potential, micromotion and ac magnetic fields should be correlated.  As the trap is a predominately reactive load, micromotion and magnetic fields should be $\sim 90^\circ$ out of phase.  There would also be a spatial correlation but this would likely have a rather complex dependence on design and heavily dependent on fabrication imperfections.  However, it may still be possible to mitigate these effects by design, particularly as ion traps move to chip-scale fabrication technologies \cite{mehta2014ion, mehta2016integrated}.

The discussion here has been restricted to magnetic fields and is a straightforward application of the Wigner-Eckart theorem.  As the Wigner-Eckart theorem applies to any tensor operator, similar considerations should be given to other fields.  In an ion-trap system, the trapping field itself will interact with the ion through the quadrupole moment.  In this case it will induce couplings between $\Delta m=0,\pm 1, \pm 2$ states.  The treatment given in \cite{ItanoQuad} for the static case can be readily generalised and these effects will be considered in future work. Similar results to those here can be anticipated but some differences would arise.  Owing to the tensor nature of the interaction it would likely influence both linear and quadratic Zeeman shifts and energy shifts would also depend on trap geometry.  It would also contribute to an Autler-Townes splitting or rf sideband for levels supporting a quadrupole moment.
\begin{acknowledgements}
We would like to thank the ion storage group at the National Institute of Standards and Technology for fruitful discussions and bringing our attention to the work in \cite{meir2018experimental}.  We acknowledge the support of this work by the National Research Foundation, Prime Ministers Office, Singapore and the Ministry of Education, Singapore under the Research Centres of Excellence programme. This work is also supported by A*STAR SERC 2015 Public Sector Research Funding (PSF) Grant (SERC Project No: 1521200080) and the Ministry of Education, Singapore, under the Education Academic Research Fund Tier 2 grant (Grant No.  MOE2016-T2-1-141). T. R. Tan acknowledges support from the Lee Kuan Yew postdoctoral fellowship.
\end{acknowledgements}
\bibliography{MagneticFields}
\bibliographystyle{unsrt}
\end{document}